\documentclass{moriond}

\def\be{\begin{equation}}
\def\ee{\end{equation}}
\def\bea{\begin{eqnarray}}
\def\eea{\end{eqnarray}}

\usepackage{amsmath}  
\usepackage{enumitem}

\newcommand{\Photo}{}

\begin{document}
\vspace*{4cm}
\title{Multimessenger emission from choked jets in collapsar}

\author{A. Zegarelli$^{1,*}$, M. Pais$^{2}$, E. Peretti$^{3}$, S. Celli$^{4,5}$}
\footnotetext{* Speaker: angela.zegarelli@astro.ruhr-uni-bochum.de}

\address{\footnotesize
$^{1}$ AIRUB, Ruhr University Bochum, Germany; 
$^{2}$ INAF Padova, Italy; 
$^{3}$ INAF Arcetri, Italy; 
$^{4}$ Sapienza Università di Roma, Italy; 
$^{5}$ INFN Roma, Italy
}

\maketitle\abstracts{
The death of massive stars produces central accreting compact objects and sometimes relativistic jets. Not all jets escape the stellar envelope: unsuccessful, or choked, jets dissipate their energy into a pressurized cocoon, which expands and may break out as a mildly relativistic outflow. We investigate the plasma physics of collapsing massive stars hosting choked jets through relativistic, non-resistive magnetohydrodynamical simulations. We delineate the parameter space for jet choking and quantify the acceleration rate and efficiency of charged particles at strong shocks, which are potential sources of high-energy neutrinos and electromagnetic transients. Our study focuses on blue and red supergiant progenitors, both promising candidates for jet choking.}

\vspace{-0.2cm}
\section{Introduction}
\vspace{-2mm}
The death of massive stars leads to core-collapse supernovae, forming central compact objects and sometimes launching relativistic jets~\cite{gottlieb_black_2022}. Jets from hydrogen-stripped progenitors can escape and generate long Gamma-Ray Burst (GRBs), while those in extended stars, such as Red Super Giants (RSGs) and Blue Super Giants (BSGs), may stall, producing choked GRBs~\cite{Piran2019}~\cite{bromberg_propagation_2011}.  
Even choked jets inflate a high-pressure cocoon that can break out as a mildly relativistic outflow, producing UV/X-ray transients. Choked GRBs are also promising hidden sources of high-energy neutrinos, as the dense stellar environment enables hadronic interactions before jet termination~\cite{murase2016}~\cite{He2018}~\cite{fasano2021}. 
Such hidden sources naturally avoid the tension between observed neutrinos and the lack of corresponding gamma rays~\cite{Ackermann2015}, and may contribute to the unresolved diffuse flux detected by IceCube~\cite{icecube_diffuse}.
Previous studies mostly relied on analytic treatments. Here we perform relativistic magnetohydrodynamical (RMHD) simulations including jet magnetization to study jet propagation, shocks, and cocoon dynamics. This approach provides a more accurate evaluation of particle acceleration and the prospects for electromagnetic (EM) and neutrino detection. We focus on extended progenitors (BSGs and RSGs), where large radii and dense envelopes hinder jet propagation and favor jet choking, strongly affecting EM and neutrino signatures.

\vspace{-0.3cm}
\section{Hydrodynamical evolution of jet's motion in static external media}
\label{sec: setup}
\vspace{-0.3cm}
\subsection{Stellar setup}
\label{subsec: stellar_setup}
\vspace{-2mm}
The stellar structure is modeled with a continuous polytropic density profile:
\begin{equation}
\label{eq: stellar_density_profile}
    \rho(r) = \rho_0 \left( \frac{r}{R_*} \right)^{-\alpha} \left( 1- \frac{r}{R_*} \right)^n \,,
\end{equation}
where $R_*$ and $M_*$ are the stellar radius and mass, $\alpha$ controls the inner slope, and $n$ the outer envelope. The normalization $\rho_0$ is set so that the profile integrates to $M_*$.   For BSGs, we adopt a single polytrope with $\alpha=2.5$ and $n=3$, while for RSGs we combine two polytropes: one for the core ($\alpha=2.5$, $n=3$) and one for the extended envelope ($\alpha=1.5$, $n=1.5$). As benchmark values, for BSGs we adopt $M_*=15\,M_\odot$ and $R_*=50\,R_\odot$; while, for RSGs we adopt $M_\mathrm{core}=3\,M_\odot$, $M_\mathrm{env}=12\,M_\odot$ (so that $M_*=M_\mathrm{core}+M_\mathrm{env}$), with radii $R_\mathrm{core}=3\,R_\odot$ and $R_\mathrm{env}=R_*=500\,R_\odot$. 
\vspace{-0.3cm}
\subsection{Breakout and choking times}
\label{subsec: numerical_estimations}
\vspace{-2mm}
A simple criterion to determine whether a jet successfully drills through the star or becomes choked is given by the integral equation~\cite{pais_velocity_2023}:
\begin{equation}
\label{eq: eqmotion}
t - \int_0^t \beta_h(t') \, \mathrm{d}t' = t - \frac{r_h(t)}{c} = t_\mathrm{eng} \,,
\end{equation}
where \(t_\mathrm{eng}\) is the engine duration, and \(r_h(t)\) and \(\beta_h(t)\) denote the position and velocity of the \textit{jet head}, i.e., the forward shock at the head of the jet. The \textit{jet tail} is the last element launched by the engine, which propagates nearly at the speed of light. When the tail and head become causally connected, the jet–cocoon system propagates adiabatically along the decreasing density gradient. Using these arguments, before running the full RMHD simulations, we estimated breakout and choking times by numerically integrating Eq.~\eqref{eq: eqmotion}. The integration was initialized with a jet radius \(r_j = 10^4~\mathrm{km}\) and a cocoon radius \(r_\mathrm{c} = 0\), neglecting general relativistic effects from the central compact object. We varied the jet parameters (\(L_\mathrm{jet}, \theta_\mathrm{jet}\)), where \(L_\mathrm{jet}\) denotes the jet luminosity and \(\theta_\mathrm{jet}\) its initial opening angle, over thousands of 1D simulations. Choking and breakout times were determined by root-finding the instant when the jet tail catches the jet head. From these simulations, we derive the following relations for the choking time:
\begin{equation}
\label{eq:tch}
t_{\rm ch}~[\rm s] \simeq 10\left(\frac{t_{\rm eng}}{10\,{\rm s}}\right) +
\begin{cases}
39\left(\frac{t_{\rm eng}}{10\,{\rm s}}\right)^{1.27}
\left(\frac{L_{\rm jet}}{10^{51}\,{\rm erg\,s^{-1}}}\right)^{0.34}
\left(\frac{\theta_{\rm jet}}{5^\circ}\right)^{-1.41}, & \text{BSG} \\[1mm]
60.3\left(\frac{t_{\rm eng}}{10\,{\rm s}}\right)^{0.875}
\left(\frac{L_{\rm jet}}{10^{51}\,{\rm erg\,s^{-1}}}\right)^{0.205}
\left(\frac{\theta_{\rm jet}}{5^\circ}\right)^{-0.874}, & \text{RSG}
\end{cases}
\end{equation}
and for the breakout time:
\begin{equation}
\label{eq:tbo}
t_{\rm bo}~[\rm s] \simeq 
\begin{cases}
116\left(\frac{R_*}{50\,R_\odot}\right)
+ 18\left(\frac{L_{\rm jet}}{10^{51}\,{\rm erg\,s^{-1}}}\right)^{-0.24}
\left(\frac{\theta_{\rm jet}}{5^\circ}\right)^{0.94}, & \text{BSG} \\[1mm]
1160\left(\frac{R_*}{500\,R_\odot}\right)
+ 195\left(\frac{L_\mathrm{jet}}{10^{51}\,{\rm erg\,s^{-1}}}\right)^{-0.227}
\left(\frac{\theta_\mathrm{jet}}{5^\circ}\right)^{0.925}, & \text{RSG}
\end{cases}
\end{equation}
\begin{figure*}[t!]
\centering
\includegraphics[width=1.0\linewidth, trim=10 10 10 10, clip]{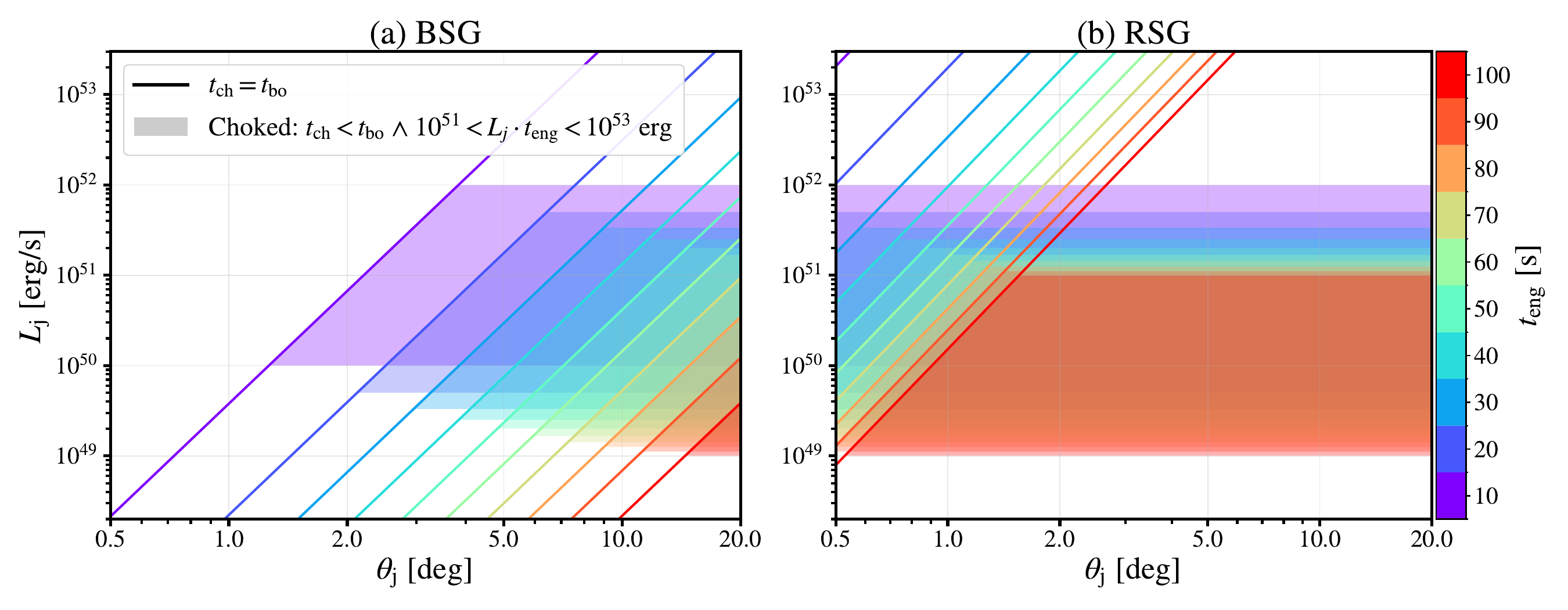}
\caption{Choked-jet parameter space in the $(L_\mathrm{jet}, \theta_\mathrm{jet})$ plane for the representative (a) BSG and (b) RSG progenitors.   Solid curves indicate the condition $t_\mathrm{ch} = t_\mathrm{bo}$; below these curves, the jet remains choked within the stellar envelope. Shaded regions show the range of jet energies satisfying $10^{51}~\mathrm{erg} < L_\mathrm{jet}\, t_\mathrm{eng} < 10^{53}~\mathrm{erg}$.}
\label{fig:param_space_combined}
\end{figure*}
These analytical estimates for $t_{\rm ch}$ and $t_{\rm bo}$, valid for the BSG and RSG masses and radii considered here (Sec.~\ref{subsec: stellar_setup}), provide a simple and efficient way to explore the jet--star interaction parameter space. They indicate when a relativistic, unmagnetized jet remains choked or successfully escapes to produce a GRB, and are used to set the initial conditions for the RMHD simulations. This approach reduces computational cost while remaining complementary to the higher accuracy of the RMHD results. Figure~\ref{fig:param_space_combined} shows the conditions under which jets with varying engine durations \(t_\mathrm{eng}\) remain choked within the stellar envelopes of BSG and RSG progenitors. The solid curves mark the locus where \(t_\mathrm{ch} = t_\mathrm{bo}\); jets below these curves fail to break out. Shaded regions indicate the energetically viable parameter space, defined by 
\(
10^{51}~\mathrm{erg} < L_\mathrm{jet}\, t_\mathrm{eng} < 10^{53}~\mathrm{erg}
\), 
with the left and right panels corresponding to BSG and RSG progenitors, respectively.
The figure illustrates that RSGs allow a substantially larger choked-jet parameter space than BSGs. They can accommodate more collimated and more luminous jets for the same fiducial luminosity \(L_\mathrm{jet} = 10^{51}~\mathrm{erg/s}\) and opening angle \(\theta_\mathrm{jet} = 5^\circ\). In general, longer engine durations require lower jet luminosities and wider opening angles for the jet to remain choked.
\vspace{-0.3cm}
\section{Non-radiative RMHD simulations}
\vspace{-0.3cm}
\subsection{Computational setup}
\label{subsec: computational_setup}
\vspace{-2mm}
We perform our simulations using the parallel multidimensional RMHD code {\textsc{pluto}} (v4.4.3)\footnote{Latest version available during this work}~\cite{mignone_pluto_2007}, which integrates conservation laws on a rectangular grid using a finite-volume, shock-capturing scheme. We employ the special relativistic hydrodynamics module in 2.5D spherical coordinates ($r,\theta$, with $v_\phi$ and $B_\phi$ evolved with periodic boundary conditions), a Taub-Matthews equation of state~\cite{mignone_equation_2007} with $\gamma=5/3$ for non-relativistic and $\gamma=4/3$ for relativistic flows, and a divergence cleaning scheme for magnetic fields~\cite{dedner_hyperbolic_2002}.
\vspace{-0.3cm}
\subsection{Simulation conditions}
\vspace{-2mm}
The initial conditions for our preliminary RMHD simulations of RSG and BSG progenitors were conservatively selected to ensure jet choking in a magnetized scenario, using Eqs.~\eqref{eq:tbo} and \eqref{eq:tch} as upper limits. For the results presented here, we adopt a reference engine duration of $t_\mathrm{eng} = 5~\mathrm{s}$ and a jet luminosity of $L_\mathrm{jet} = 10^{50}~\mathrm{erg\,s^{-1}}$. The significant difference in stellar radii between the two progenitor classes requires distinct numerical resolutions and integration strategies for the jet propagation equations. Here we focus on a BSG model under these reference conditions, while simulations for RSG progenitors are ongoing and currently under validation.
\vspace{-0.5cm}
\subsection{Shock finder}
\vspace{-2mm}
Shock regions play a key role in our study, as they are the primary sites of particle acceleration and therefore potential sources of high-energy neutrinos and EM emission. Identifying and characterizing shocks in the jet--cocoon system is thus essential to connect the RMHD dynamics with observable multimessenger signatures. To locate such regions in the simulations, we flag grid cells as shocks if they simultaneously satisfy:
\begin{itemize}[noitemsep, topsep=0pt, leftmargin=*]
    \item Converging flows ($\nabla \cdot \vec{v} < 0$), indicating compression;
    \item Significant pressure jump between adjacent cells;
    \item Alignment of temperature and density gradients ($\nabla T \cdot \nabla \rho > 0$), to filter out spurious detections;
    \item High Mach number within jet material, selecting strong shocks ($\mathcal{M} = v_\mathrm{shock}/v_\mathrm{sound}$).
\end{itemize}
These criteria allow us to isolate strong, physical shocks where diffusive shock acceleration is expected to operate efficiently, enabling a quantitative assessment of particle acceleration in the jet environment.

\begin{figure}[t!]
    \centering
    \includegraphics[width=1\linewidth]{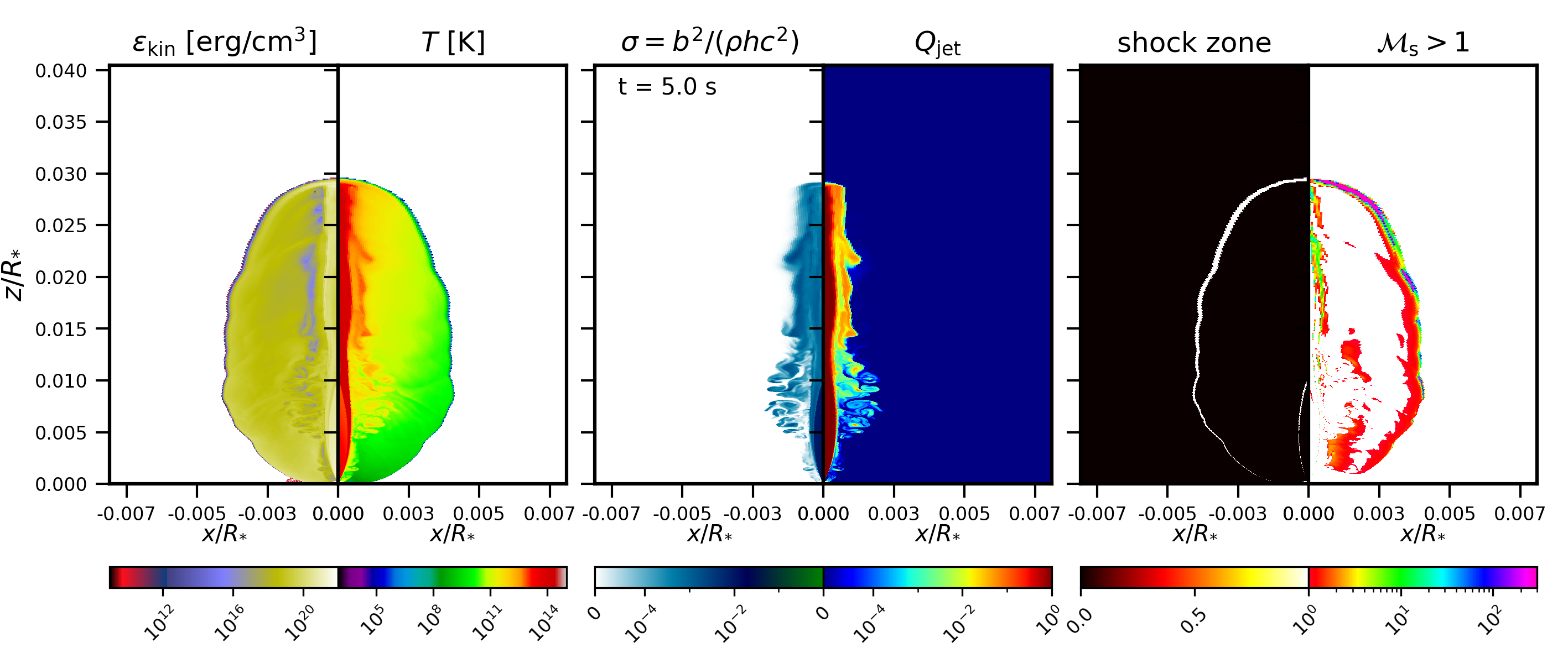}
    \includegraphics[width=1\linewidth]{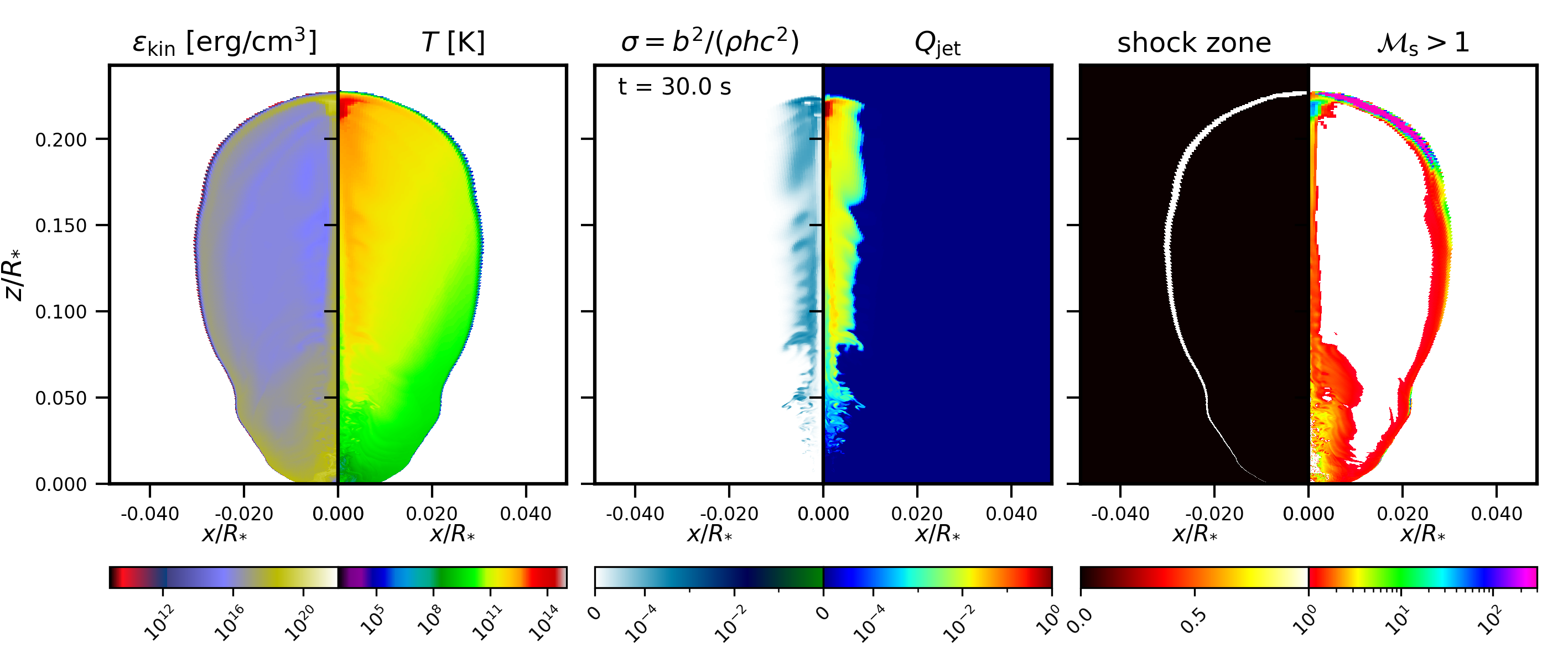}
    \caption{Snapshots from the RMHD simulation of a relativistic jet propagating through the stellar envelope of a BSG ($L_{\rm jet} = 10^{51}~\rm erg\,s^{-1}$) at two representative times. Each row shows, in order: kinetic energy density, temperature, relativistic magnetization, a scalar tracer of the jet, shock-flagged cells, and the relativistic Mach number $\mathcal{M}_\mathrm{s} \geq 1$. The plot axes are scaled to the stellar radius, and an uneven aspect ratio is used to enhance the apparent jet width. \textit{Top row}: after 5~s of engine activity. \textit{Bottom row}: after 30~s of evolution.}
    \label{fig: jet_sim}
\end{figure}

\vspace{-0.3cm}
\section{Preliminary RMHD results for a BSG progenitor}
\vspace{-2mm}
We present preliminary RMHD results for a relativistic jet propagating through a BSG progenitor. The jet is injected for $t_{\rm eng} = 5~\mathrm{s}$ and then evolved for $25~\mathrm{s}$ post-engine shutdown. Snapshots at $t = 5~\mathrm{s}$ and $t = 30~\mathrm{s}$, in Fig.~\ref{fig: jet_sim}, show the pressure, jet tracer, shock-flagged cells, and relativistic Mach number $\mathcal{M}_\mathrm{s} >1$. During injection, a high-pressure cocoon forms, producing the forward cocoon shock, the termination shock, and an internal collimation shock. After engine shutdown, the collimation shock fades, leaving the forward and cocoon shocks as dominant acceleration sites. These high-Mach shocks provide potential sites for particle acceleration, with emergent neutrino or electromagnetic signals expected mainly from the forward shocks in the post-injection phase. Simulations for RSG progenitors are ongoing and will be presented in a forthcoming paper (\textit{in prep.}~\cite{pais_neutrino_radiation_prep_2025}).

\vspace{-0.6cm}
\begingroup
\section*{References}
\footnotesize
\vspace{-0.34cm}
\bibliography{moriond}
\endgroup

\end{document}